\documentclass[a4paper,10pt]{iopart}
\usepackage[english]{babel}
\usepackage[utf8]{inputenc}
\usepackage[T1]{fontenc}
\usepackage{amssymb}
\usepackage{epsf}
\usepackage{color}
\usepackage{colordvi}
\usepackage{graphicx}
\usepackage{textcomp}
\usepackage{cite}
\usepackage{textcomp}
\usepackage{ulem}
\begin{document}
\title{Identifying the mechanism of single-domain single-layer MoS$_2$ growth on Au(111)}
\author{Moritz {Ewert}$^{1,2,3}$, Lars {Bu\ss}$^{1,2}$, Jeppe V. Lauritsen$^4$, Jens {Falta}$^{2,3}$ and Jan Ingo {Flege}$^{1,2,3}$}
\address{$^1$ Applied Physics and Semiconductor Spectroscopy, Brandenburg University of Technology Cottbus-Senftenberg, Cottbus, Germany}
\address{$^2$ Institute of Solid State Physics, University of Bremen, Bremen, Germany}
\address{$^3$ MAPEX Center for Materials and Processes, University of Bremen, Bremen, Germany}
\address{$^4$ iNANO - Interdisciplinary Nanoscience Center, Aarhus University, Aarhus, Denmark}
\ead{flege@b-tu.de}
\begin{abstract}
The nucleation and growth of single-layer molybdenum disulfide single domain islands is investigated by \textit{in situ} low-energy electron microscopy. We study the growth of micron-sized flakes and the correlated flattening process of the gold surface for three different elevated temperatures. Furthermore, the influence of surface step edges on the molybdenum disulfide growth process is revealed. We show that both island and underlying terrace grow simultaneously by pushing the surface step in the expansion process. Our findings point to an optimized growth procedure allowing for step-free single-domain single-layer islands of several micrometers in size, which is likely transferable to other transition metal dichalcogenides.
\end{abstract}
\maketitle
\section{Introduction}
Since the first studies illustrating the extraordinary electronic properties of graphene \cite{Novoselov2004Science}
, similar characteristics were predicted for other 2D materials, such as single-layer (SL) transition metal dichalcogenides (TMDs)\cite{Novoselov2005PNAS}.
A promising example in terms of applications and probably the most prototypical one is SL molybdenum disulfide (MoS$_2$) as the band structure has been predicted to exhibit a direct band gap \cite{Lebegue2009PRB}.
Indeed, this transition from an indirect semiconductor in its bulk form to a direct semiconductor as a 2D SL has also been demonstrated experimentally \cite{Mak2010PRL, Splendiani2010NanoLett},  
 resulting in attractive optoelectronic properties\cite{Ramasubramaniam2012PRB}
  and an interesting valley/spin valley structure\cite{Cao2012NatComm, Zeng2012NatNano}. 
First functioning MoS$_2$ transistor devices have been reported\cite{Radisavljevic2011NatMat} as well as optoelectronic devices\cite{Lopez-Sanchez2013NatNano}. Furthermore, from a chemical point of view, MoS$_2$ has since long been known to be an excellent catalyst for sulfurization and desulfurization processes \cite{Helveg2000PRL, Lauritsen2004JCatHydrodesulfurization, Bruix2015ACSNano}, rendering MoS$_2$ a versatile material for various applications in heterogeneous catalysis and chemical sensing.

Most studies have used mechanical exfoliation to produce SL MoS$_2$ \cite{Mak2010PRL,Splendiani2010NanoLett,Zeng2012NatNano,Cao2012NatComm,Radisavljevic2011NatMat,Lopez-Sanchez2013NatNano} 
whereas the research employing more industrially suitable growth methods such as chemical vapor deposition (CVD)\cite{Shi2012NanoLett, Lee2012AdvMat, Zhan2012Small, Liu2012NanoLett, Marzari2015ACSNano}
are gaining increasing interest due to its scalability for large-scale fabrication. Even though remarkable first results have been reported using CVD, a deeper understanding of the growth mechanism would further improve the growth control and hence the structural quality of the films, i.e., its single-crystalline nature. To achieve this ambitious goal, key ingredients are the suppression of defects as well as the manipulation of the island expansion to ensure the desired quality and size of the SL MoS$_2$. This knowledge is usually best obtained in dedicated epitaxial growth studies using \textit{in situ} experimental methodology for a model system, here MoS$_2$ on Au(111).

So far, epitaxial growth studies of this MoS$_2$/Au(111) model system have focused on investigating the electronic structure and catalytic properties, mostly addressing individual SL nanoclusters and islands found at low coverage\cite{Lauritsen2007NatNano, Bruix2016Faraday} 
as well as monolayer-like coverages of many SL MoS$_2$ islands \cite{Miwa2015PRL, Sorensen2014ACSNano, Gronborg2015Langmuir}. Whereas bulk MoS$_2$ may occur in various polymorphs, SL MoS$_2$ has been found to exist in its thermodynamically most stable form, the three-fold symmetric 1H-structure, where the islands are terminated by S-edges and Mo-edges at high and low values of the sulfur chemical potential, respectively\cite{Cao2015JPhysChemC}.
The six-fold symmetry of the first layer of the Au substrate facilitates the growth of MoS$_2$ mirror domains, i.\,e., of islands azimuthally rotated by 60$^\circ$ \cite{Bruix2016Faraday, Gronborg2018NatComm}. Recently, Bana \etal \cite{Bana20182DMater} presented a growth study where they were able to suppress one of the mirror domains and get large single-domain SL MoS$_2$, whereas a study by Bignardi \etal \cite{Bignardi2019PhysRevMat} reported similar results of mostly single-domain SL WS$_2$, demonstrating the feasibility of modifying the epitaxial growth process by a suitable choice of parameters. This single-domain coverage enabled revealing the spin structure of WS$_2$ by Eickholt \etal \cite{Eickholt2018PRL}, suggesting that a precise control of the growth process is highly desirable for large-scale integration of SL TMD and the exploitation of their superior materials properties in applications. The origin of the observed asymmetry, however, remained ambiguous, leaving a knowledge gap in our understanding of the growth process of MoS$_2$, and especially the transition from nanoclusters to large single-domain SL.

Here, we present the first \textit{in situ} study in which we finally unravel the mechanism that is fostering the growth of a single orientation of MoS$_2$ islands on the Au surface. Following the nucleation and growth process in real-time allows us, for the first time, to reveal and explicitly focus on the dynamic interplay between the substrate and the occurring islands leading to this distinct domain distribution. Furthermore, our results strongly indicate that the proposed mechanism very likely is a key feature promoting single-domain growth of TMDs on Au(111).

\section{Experimental}
Au(111) single crystal substrates were cleaned by Ar$^+$ ion sputtering at 0.5 kV for 1 h followed by thermal annealing at 700 $^\circ$C for 30 min. This procedure was repeated several times before the cleaning process was finalized by annealing at 950 $^\circ$C for 5 min. The deposition experiments started with the surface showing a well defined herringbone reconstruction in low-energy electron diffraction (LEED) as reported by Barth et al. \cite{Barth1990PRB} for a clean Au(111) surface.

The in-situ characterization of MoS$_2$ growth on Au(111) was performed using a commercial ELMITEC LEEM III microscope under ultra-high vacuum (UHV) conditions with a base pressure of 1$\times$10$^{-10}$ torr. Low-energy electron microscopy (LEEM) allows to directly monitor the sample surface at video rate with a spatial resolution of about or even below 10 nm. Being based on diffraction, LEEM enables recording either real space (imaging mode) or reciprocal space images (diffraction mode).

LEEM also allows for choosing among different contrast modes. A contrast aperture placed into the back focal plane of the objective lens (where the diffraction pattern is generated) can be used to select an imaging beam and suppress the others. The (00) specular beam is used for bright-field imaging, which is applied here for observing MoS$_2$ growth in time-resolved imaging sequences. In addition, by selecting a particular non-specular beam in so-called dark-field imaging, the resulting diffraction contrast can be used to discriminate between different phases (here rotational domains of MoS$_2$/Au(111)) since only areas on the sample surface will appear bright in the image if they contribute to the intensity of the particular diffracted beam in reciprocal space.

Inserting an illumination aperture into the incident beam, the illuminated area of the sample can be constrained down to 250 nm, facilitating the recording of diffraction patterns of local origin. This diffraction mode is called  \textmu LEED. Micro-spectroscopy and spectro-microscopy can be performed with LEEM by spatially resolved recording of the electron reflectivity as a function of electron energy. Due to its analogy to the well-established method of I(V)-LEED\cite{Pendry1974}, this mode is called I(V)-LEEM. The corresponding experimental procedure is based on measuring the electron reflectivity in bright-field LEEM imaging mode for a range of electron energies, typically from the vacuum level up to a few tens of electronvolts. The corresponding energy-dependent reflectivity curve can be extracted with pixel resolution, yielding a lateral spatial resolution near or even below 10 nm in favorable circumstances. The shape of the curve is derived from the band structure of the unoccupied states of the sample near-surface region, resulting in a material-specific and chemically sensitive structural fingerprint \cite{Flege2014PSSRRL}. Further information on LEEM methodology can be found in the literature\cite{Bauer2014SurfMicroscopyLEE, Flege2012CharMat}.

For MoS$_2$ growth, sulfur was supplied from a dimethyldisulfide (DMDS) vapor source  backfilling the UHV chamber to 1$\times$10$^{-6}$ torr when dosing. DMDS is a suitable and safe sulfiding agent since it was previously shown to lead to facile formation of single-layer MoS$_2$ on Au(111) in the temperature range of interest here \cite{Tuxen2011JCat}. The DMDS vapor background pressure was upheld during the entire growth process reported here. Molybdenum (purity 99.95\%) was evaporated from a home-built electron beam evaporator similar to existing commercial designs. MoS$_2$ growth was performed in an experiment series employing three different substrate temperatures, i.\,e., 650 $^\circ$C, 700 $^\circ$C, and 750 $^\circ$C. To ensure sufficiently stable imaging conditions, the sample was heated to the desired temperature by electron beam heating 10 min prior to the growth process. Thereafter, molybdenum was deposited at a very low rate of 0.108ML/h onto the sulfur-terminated sample surface. The growth rate has been determined from the overall coverage of MoS$_2$ islands on the Au(111) surface extracted from large-scale LEEM survey images and set into relation to the total evaporation time of Mo.

\section{Results and Discussion}
All deposition experiments started with exposing the Au(111) surface at growth temperature to DMDS (no Mo flux) for 10 min.  For the lower growth temperatures of 650 $^\circ$C and 700 $^\circ$C, this procedure resulted in lifting the herringbone reconstruction of the clean Au(111) surface and forming a (1$\times$1) LEED pattern. This is attributed to sulfur deposition on the Au(111) surface \cite{Liu2002SurfSci}. For the highest growth temperature of 750 $^\circ$C, however, the herringbone reconstruction was visible in diffraction mode after the deposition, pointing towards a desorption of sulfur atoms at this temperature.

MoS$_2$ nucleation is initiated by reactive deposition of Mo on the sample surface in the DMDS background pressure. Throughout the deposition, the sample is held at the desired growth temperature, i.\,e., 650 $^\circ$C, 700 $^\circ$C or 750 $^\circ$C. After deposition of the first few mML (0.001 monolayers), island nucleation was observed in the LEEM image, reflecting that the critical nucleus size had been reached under these conditions. In all observed cases, the nucleation centers of the islands are found on the Au terraces; no tendency for preferential nucleation at steps was observed.

\begin{figure}[tbp]
\includegraphics[width=\columnwidth]{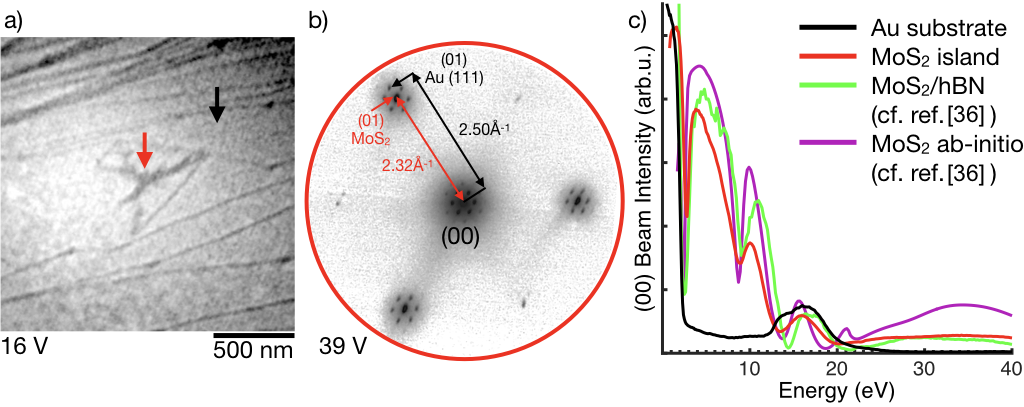}
\caption{(a) LEEM analysis at an initial stage of growth after deposition of 0.009 ML MoS$_2$/Au(111) at a substrate temperature of 650 $^\circ$C:
(a) LEEM image recorded at an electron energy of 16 eV;  at this energy the substrate appears bright and the MoS$_2$ is dark. The fine lines on the substrate surface show the step edges.
(b) \textmu LEED  of a MoS$_2$-rich area at an electron energy of 39 eV using a 1 \textmu m illumination aperture (contrast inverted). 
The typical moir\'e pattern of MoS$_2$/Au(111) is visible around the integer spots of the Au substrate. The reciprocal distances are highlighted by arrows.
(c) I(V)-LEEM spectroscopy curves of the substrate (black) and an island (red) as indicated in a).
The I(V)-LEEM curves identify the two different materials as Au(111) and SL MoS$_2$.
For comparison, experimental reference data of exfoliated SL MoS$_2$ on hBN (green line) as well as an I(V) curve of SL MoS$_2$ calculated using \textit{ab initio} scattering theory (violet line) have been added \cite{DeJong2018pssb}.
}
\label{fig:initialgrowth1}
\end{figure}

In fig.\,\ref{fig:initialgrowth1}a) a newly nucleated MoS$_2$ island (growth temperature: 650 $^\circ$C) is shown. The corresponding \textmu LEED diffraction pattern (fig.\,\ref{fig:initialgrowth1}b)) of a single MoS$_2$ island has been recorded from a MoS$_2$ covered area of 1 \textmu m in diameter after the growth. It is clearly dominated by the moiré structure of MoS$_2$/Au(111), which will be analyzed below. The diffraction pattern is threefold symmetric, as expected for a single domain of SL MoS$_2$(0001). Furthermore, spectromicroscopy, as shown in fig.\,\ref{fig:initialgrowth1}c), clearly confirms the island to consist of a single layer of MoS$_2$.
%

From the exact position of the moir\'e spots relative to the three integer reflections, the lattice mismatch between MoS$_2$ and Au(111) and thus the MoS$_2$ lattice parameter is determined to 3.14 \AA, in excellent agreement with the in-plane lattice constant of the bulk crystal \cite{Young1968JPD}. Together with the threefold symmetry of the diffraction pattern, our findings suggest the formation of unstrained single-layer islands of 1H-MoS$_2$ on Au(111), consistent with previous reports \cite{Sorensen2014ACSNano, Gronborg2015Langmuir}.


Generally, the crystalline and electronic structure of any material determine the specular electron reflectivity depending on kinetic electron energy, the so-called $I(V)$ curve, allowing to locally identify a material and its structure by comparing the recorded $I(V)$ curves to calculated spectra or experimental reference data \cite{Flege2014PSSRRL}. Here, after the growth we have identified only two distinct types of $I(V)$ curves representing the substrate and the islands as shown in fig.\,\ref{fig:initialgrowth1}c). The $I(V)$ fingerprint of the island (fig.\,\ref{fig:initialgrowth1}c)) is in good agreement with the $I(V)$ reference curve of exfoliated single-layer MoS$_2$ on hBN presented by DeJong \etal \cite{DeJong2018pssb} and clearly differs from the corresponding bi- and trilayer $I(V)$ curves (not shown). Strikingly, an even better match is observed when comparing our $I(V)$ curve of MoS$_2$/Au(111) to the calculated $I(V)$ spectra for SL MoS$_2$ as presented in \cite{DeJong2018pssb} based on ab initio scattering theory \cite{Krasovskii_PhysRevB_2004}. This finding implies that the orbitals facilitating the bonding to the substrate do neither significantly affect the bandstructure above the vacuum level nor do they facilitate an efficient coupling to the incident plane wave in the scattering process. It is, however, noted that angle-resolved photoemission (ARPES)  and density functional theory calculations have shown that substrate bonding and modification of the out-of-plane orbitals associated with the $\Gamma$-point is relevant for MoS$_2$ bonded to Au(111) \cite{Bruix2016Faraday,Miwa2015PRL}. Taken together, these observations suggest that the bonding orbitals at the interface are indeed too strongly localized to effectively couple to the incident plane wave, rendering them invisible to $I(V)$ microspectroscopy.

\begin{figure}[tbp]
\includegraphics[width=\columnwidth]{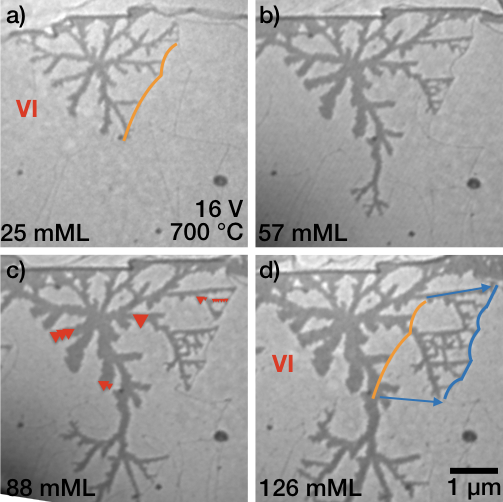}
\caption{LEEM images during step pushing growth of a selected MoS$_2$ island at a growth temperature of 700 $^\circ$C recorded at an electron energy of 16 eV at a deposit of a) 25 mML, b) 27 mML, c) 88 mML, d)125 mML. The triangular shape of the MoS$_2$ is highlighted by red triangles in c). The orange line indicates the location of the step bunch in (a); it is also shown in (d) together with its current position (blue line). The progression of this step bunch is indicated by arrows. In the area marked VI the island has stopped ramification growth. We focus on this behavior in fig.\,\ref{fig:stepcontact}.
}
\label{fig:Growth700C}
\end{figure}

At a later stage of the deposition at all selected temperatures, the islands growth proceeds by ramification as depicted in the LEEM time-lapse sequence in fig.\,\ref{fig:Growth700C} for growth of MoS$_2$ at 700 $^\circ$C. At this electron energy (16 eV) the bare Au substrate appears bright and the MoS$_2$ islands are dark. Furthermore, the step edges of the Au(111) surface appear dark in the LEEM images due to phase contrast.

In the growth regime of strong excess sulfur supply as applied here, literature suggests that MoS$_2$ islands are fully sulfided and that preferentially single-layer MoS$_2$ islands with a triangular morphology are formed.\cite{Cao2015JPhysChemC, Bruix2016Faraday, Gronborg2018NatComm}. The predominant triangular morphology results from a higher stability of one type of zig-zag edge termination of a single layer MoS$_2$ sheet, the so-called (10$\overline{1}$0) Mo-edge, as compared to the other, i.\,e., the ($\overline{1}$010) S-edge. Most isolated islands therefore assume either a perfect triangular shape or a shape of a slightly truncated triangle with long Mo-edges and short S-edges. This is also the case for the MoS$_2$ structure in fig.\,\ref{fig:Growth700C}, where the expanding island keeps its three-fold symmetry as it grows in substructures of the typical triangular shape out of the extending branches (see red triangles in fig.\,\ref{fig:Growth700C}c)). As pointed out by Gr\o nborg \etal \cite{Gronborg2015Langmuir}, there are two rotational domains of epitaxially grown MoS$_2$ on Au(111) due to the six-fold symmetry of the first substrate layer and the three-fold symmetry of MoS$_2$. In fig.\,\ref{fig:Growth700C}, all the triangular substructures point in the same direction, reflecting that the ramified island consists of a coherent single rotational domain of MoS$_2$.

From the analysis of an expanding growth front, it is also observed that whenever the island reaches a substrate step edge, the island either stops expanding in this direction, or makes the step edge move itself, thereby expanding the Au terrace beneath the island. The latter case is illustrated in fig.\,\ref{fig:Growth700C}. At growth stage a) (i.e. 25 mML) a downward step bunch on Au(111) is present at the position identified by an orange line. During growth the step bunch moves in front of the growing island, as seen at growth stages b) and c) (i.e. 57 and 88 mML). This process is ongoing over the entire growth sequence analyzed here. The progression of the step bunch from 25 to 126 mML is indicated by the orange and blue lines as well as the arrows in fig.\,\ref{fig:Growth700C}(d). 

\begin{figure}[tbp]
\includegraphics[width=\columnwidth]{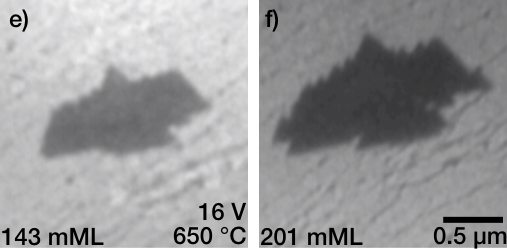}
\caption{LEEM images during growth of a minority domain MoS$_2$ island at a growth temperature of 6500 $^\circ$C recorded at a electron energy of 16 eV for a deposit of 
(a) 80 mML and (b) 115 mML.
The island keeps a compact triangular edge; neither ramification nor step pushing is observed. 
}
\label{fig:Growth650C}
\end{figure}

Interestingly, such step movement (referred to as step-pushing in the following) can only be observed for growth of the MoS$_2$ majority domain. For the MoS$_2$ minority domain, no indication of step-pushing is found as exemplarily shown for an island during growth at 650 $^\circ$C (figs.\,\ref{fig:Growth650C} a) and b)). Distinct from the strongly ramified shape of the majority islands, this minority island exhibits a dense, compact triangular shape without any visible branches. Obviously, the inability to push the step edge seems to limit the growth to the existing terrace on the Au surface only, in stark contrast to the previous finding of continuing growth of islands of the majority domain.

%

%

\begin{figure}[tbp]
\includegraphics[width=\columnwidth]{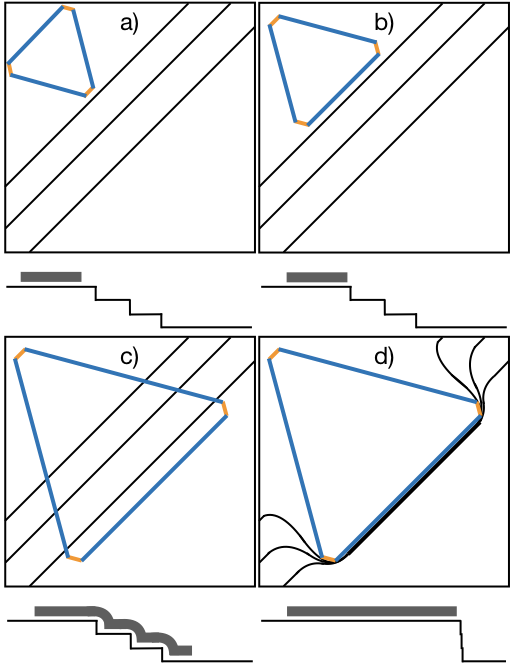}
\caption{Schematic top and side view of growth scenarios:
a) S-edge (orange line) reaches step edge,
b) Mo-edge (blue line) reaches step edge; further expansion leads to either
c) static step edges the island by overgrowth of several terraces, or
d) MoS$_2$ island growth on a single terrace by step movement and terrace expansion (without island bending).
}
\label{fig:Growthscheme}
\end{figure}

To explain the observation of majority-domain ramification growth by Au step-edge displacement in specific directions only, we note that a single domain orientation of MoS$_2$ implies two possible orientations of the triangular MoS$_2$ island relative to a [111]-oriented Au step edge. These two scenarios are schematically shown in fig.\,\ref{fig:Growthscheme}.

In scenario a), the ($\overline{1}010$) S-edge (indicated as orange lines in fig.\,\ref{fig:Growthscheme}) is parallel to the Au step edge, so the apex of the triangular island contacts the step edge upon growth. Oppositely, in scenario b), the ($10\overline{1}0$) a long side of the MoS2$_2$  island touches the step edge forming a long contact line with it. As mentioned above, this side ist the Mo-edge of the MoS2$_2$ structure (indicated as blue lines in fig.\,\ref{fig:Growthscheme}).

Our LEEM observations exclusively find a termination of growth if the apex of the island collides wit a step. , whereas the approach of a long side of an island, i.e. its  Mo-edge terminated side) initiates  two scenarios of progressive growth in two different ways, c.f. scenario overgrowth of the step or step pushing as displayed in fig.\,\ref{fig:Growthscheme} c) and d).

Scenario c) shows the MoS$_2$ island growing over several step edges without any movement of the step edges themselves.
In this scenraio the island is bent similar to a carpet to cover the height differences of the underlying substrate.
In scenario d) the movement of the step edge and consequently the growth of the terrace (step pushing) allows the island to expand further on the same terrace without any bending.

Step restructuring driven by the formation of Lander islands has already been reported for the Cu(110) \cite{Rosei2002Sci}. For this system, the formation of Cu(110) nanostructures at step edges has been found to be energetically favored by the adsorption of a single Lander molecule fitting perectly onto the metal nanostructure.
For the Cu(111) surface a process of step bunching has been found upon exfoliation and mild annealing of single layer graphene onto a foil of Cu(111) \cite{Yi2018PRL}.
The argument for massive step rearrangement found here, is an energy gain for the adsorbed graphene by the reduction of graphene bending across step edges.
A similar step-restructuring argument can be claimed here due a high mobility of surface Au atoms induced by the relatively high growth temperatures in our experiments.
Here, however, step rearrangement occurs during growth as in the case of metal nanostructure formation on C(110).
Applying this argument, the following factors influencing the growth mechanism can be derived.
Firstly, a crossing of a step edge by the. expanding the MoS2$_2$ islands would lead to a sequential downward and upward bending of the MoS$_2$ island (see fig.\,\ref{fig:Growthscheme}, lower part) the resulting stress in the island itself makes this growth step energetically unfavorable.
This conclusion is supported by our extensive growth observations which did not reveal any crossing of a step edge by a MoS$_2$ island.
Secondly, the area of contact to the step edge determines whether or not the step edge will move.
Again from an energetically point of view the diffusing Au atoms already on the step edge are more likely to remain longer at a step edge than a terrace site.
Considering the additional influence of a near MoS$_2$ island which has reached the end of the terrace and would have to bend itself to reach the next terrace this is expected to be even more favorable.
If the islands reaches to step with its Mo-egde, as sketched in  fig.\,\ref{fig:Growthscheme} scenario d), the terrace can expand parallel to the step edge keeping the length of the step edge equally long.
Therefore, the surface energy is not or only slightly influenced by the shape of the moving step edge.
If, however, the islands reaches to step with its S-edge, i.e. its triangular tip, a terrace expansion would result in a triangular step line of  an uncommon shape and additional step length. 
Consequently, this results in an energy burden and a higher formation energy.

The question if step pushing occurs in upward or downward direction of the substrate can be ansewered by LEEM using routines established by 
Bauer\,\cite{Bauer1998SurfRevLett} and Kennedy \etal \cite{Kennedy2010ProcRSocA,Kennedy2011Ultramic}. Tuning the objective lens focus accoridingly,
we find step pushing occuring on downward steps exclusively for growth temperatures of  700 $^\circ$C and lower.

For our highest growth temperature (i.e. 750 $^\circ$C), however, also a very slow movement of an upward step edge fduring growth of a MoS$_2$ island could be observed.
The fact that this only was observed at 750 $^\circ$C reflects that  the Au atoms are thermally highly activated at this temperature and detach from the step edge.
 This growth scheme, however, is the result of a delicate balance between island growth speed and step edge mobility.
Additional experiments with an increased growth rate (2.36 ML/h) show that step pushing is suppressed by the supply of additional material forcing the MoS$_2$ island to grow over the step edges before the terrace has time to expand.

\begin{figure}[tbp]
\includegraphics[width=\columnwidth]{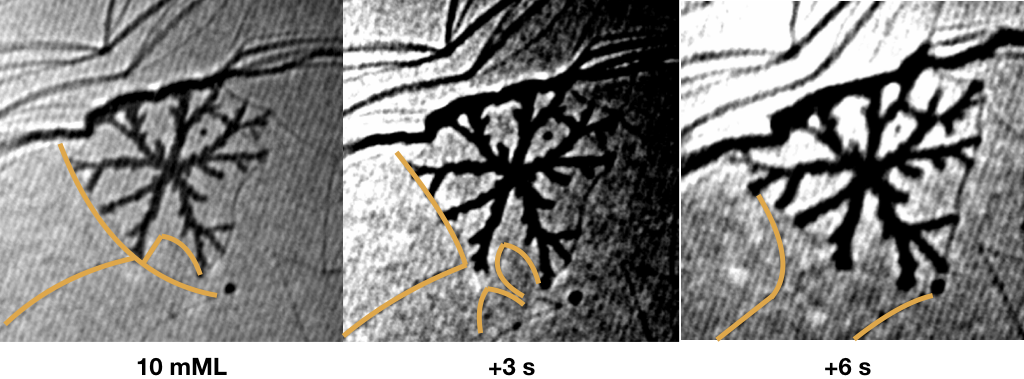}
\caption{Focus on area VI of fig.\,\ref{fig:Growth700C}.
The impact of step annihilation on growth:
LEEM images measured at an electron energy of 16 eV.
a) at the moment of step merging, b) and c) 3 and 6 seconds later, respectively.
The orange lines illustrate the position of the step edge.
}
\label{fig:stepcontact}
\end{figure}

In oder to identify the dring force for island ramnification,
a close inspection of the area in fig.\,\ref{fig:stepcontact} marked as VI is revealing.
Up to a deposit of 10mML, the island has expanded by ramnification. The underlying growth scheme
is step pushing as discussed above. 
At this point, the situation changes by the collision of two steps of identical height but opposite sign, i.e. step annihilation.
This process occures at a much faster time scale than island expansion. The collision of the steps is highlighted by orange lines
in  fig.\,\ref{fig:stepcontact}. Upon collsion, the remaining steps withdraw and leave the island branches ending on some inner part of a (newly formed) wide terrace.
Immediately, the growth scheme for these branches switches from ramnification to broadening while the branches on the other island side still in contact to a step and further progress
by ramnification. This concludes, that contact to a step edge is necessary for the ramification growth. Without step contact, the island is expanding two-dimensionally with no favored direction in the expected triangular shape.
For a substrate with a sufficiently high step density, the domination growth scheme is 
ramification growth. This favors the growth of islands with the proper alignment towards the step edges and  finally results in a high coverage for this type of islands.

\begin{figure}[tbp]
\includegraphics[width=\columnwidth]{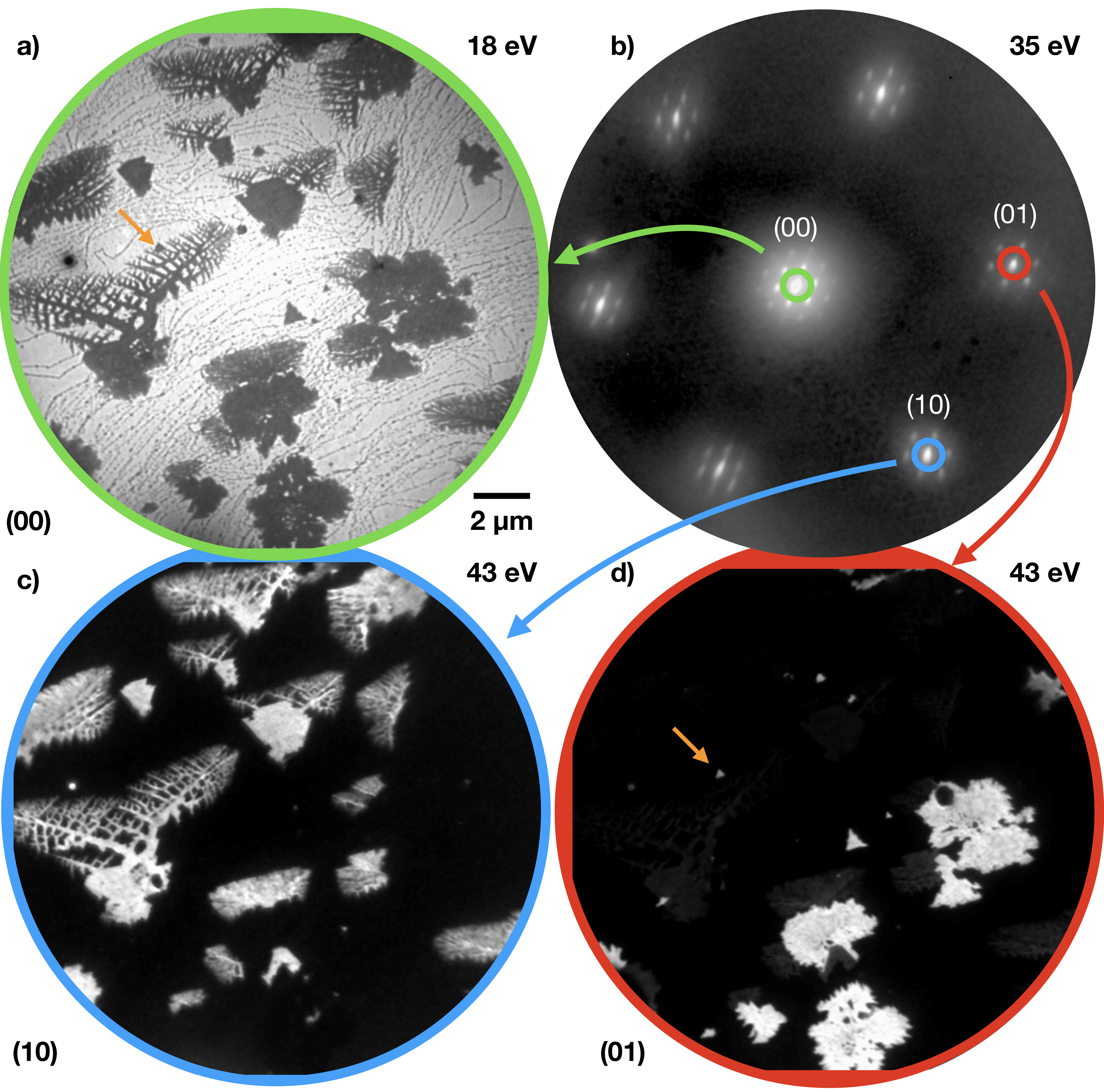}
\caption{Determination of the domain distribution of MoS$_2$ for the growth experiment at 650 $^\circ$C:
a) bright field LEEM obtained at 18 eV electron energy,
b) LEED image measured at 35. eV electron energy,
c) and d) dark-field LEEM recorded at 43 eV electron energy, using the (01) and (10) beam respectively.
The orange arrow indicates the same island in a) and d).
}
\label{fig:DFLEEM1}
\end{figure}

Depending on the miscut and its orientation, this
mechanism leads to favoring growth of one of the to existing MoS$_2$ domains over the other. 
The distribution of the two MoS$_2$ domains can be measured in LEEM using dark field imaging
as shown in fig.\,\ref{fig:DFLEEM1}.
While the two domains are indistinguishable in bright-field LEEM, the contribution of the two domains to the higher order diffraction spots varies differently with electron energy.
At 43 e, the difference between the two is large, resulting in strong contrast between the MoS$_2$ domains.
Employing the (01) beam only one of the domains appears bright in the image of fig.\,\ref{fig:DFLEEM1} c) , while imaging with the (10) beam 
provides an image of the other domain in fig.\,\ref{fig:DFLEEM1} d)  with no contribution of the first. 
The bright field image of   fig.\,\ref{fig:DFLEEM1} a) shows both domains
and is a composite of the images of fig.\,\ref{fig:DFLEEM1} c) and fig.\,\ref{fig:DFLEEM1} d).
From these images is evident, that individual islands are either consiting of the one or the other domain, i.e. they are single crystalline.
%
%
%
Inspecting large surface areas over several ten-thousand square microns for each preparation in dark field LEEM, we can determine the domain ratio for the specific growth condition. 
With increassing growth temperature, we. find an increasing disbalance.
For 650 $^\circ$C the domain ratio is determined to be 80:20, for  700 $^\circ$C we find 88:12, and for 750 $^\circ$C the. result is 90:10. 
The distribution ratio between the two possible rotational domains of MoS$_2$/Au(111) was determined by employing dark-field LEEM over several ten-thousand square microns for each preparation.

This mechanism is driven by a reduced Mo deposition rate and a high growth temperature strongly enhancing the step mobility due to surface diffusion of step atoms.
%
There is however,  an upper limit to this scheme, as with 
too high growth temperatures the mobility of Au atoms in the step edges gets so large that not only downward steps can be pushed aside by the growing MoS$_2$  islands,
but also upward steps gain mobility. At this point, the disbalance has passed its maximum.
This effect can be observed to set in at a growth temperature of 750 $^\circ$C, 
At this growth temperature, both up- and down-hill step edges can be pushed.
In the experiment both domains are found to expand over step edges, i.e. in uphill and downhill direction. 
HERE WE NEED ANOTHER FIGURE: 750 C GROWTH.
Crossing the step edges can also be triggerd by increasing the depostion rate for MoS$_2$.
Experiments with a growth rate of 2.36 ML/h exhibit an equal distribution of the MoS$_2$ domains.
We like to note, that our findings of single domain growth can be explained simply considering thermodynamic and kinetic arguments 
and do not depend on assumptions regarding the influence of the substrate registry as reported previously \cite{Bignardi2019PhysRevMat}.
\section{Conclusions}
From early days on in expitaxial growth of SL MoS$_2$ islands on Au(111) one of the main questions has been by what are the two rotational domain separated or are they equal.
With our high temperature low flux growth experiments we presented here insights showing the island orientation with respect to the local miscut is directly linked to the expansion process.
The presented mechanism of island growth of several micron size by step pushing gives a clear path on how to achieve a large single domain coverage, namely a low growth speed allowing the steps to move and enough step edges to limit the minority domain to a small initial terrace size.

This study showed at all three temperatures a dendritic growth of several micron size MoS$_2$ islands each on a single terrace.
A mechanism of step edge movement driven by optimization of free surface energy was introduced resulting in the single terrace growth.
The expansion of MoS$_2$ in single terrace growth in non-equal for the two rotational domains leading towards a single domain coverage with higher temperatures increasing the ratio of the majority domain.

This mechanism relies very much on the mobility of the surface steps.
To transfer this results to other systems one can assume that this mechanism can take place for other TMDs (e.g. WS$_2$) on the Au (111) surface.
A transfer to other substrate materials could rely on the step movement at the formation temperature of MoS$_2$ as well as the limiting temperature of the material stability.
In any case the growth rate shall be calibrated to just allow the terraces to expand in the speed of the island growth in order to achieve step-edge free single domain growth, making it readily available for photoemission spectroscopy studies into the valley and spin degrees of freedom as reported on WS$_2$/Au(111)\cite{Beyer2019PRL}.
\ack
We gratefully acknowledge support by the Central Research Fund of the University of Bremen.

 \section*{References}
\bibliographystyle{iopart-num}
\bibliography{bib/MoS2-growth.bib}

\end{document}